\documentclass[fleqn,usenatbib]{mnras}

\usepackage{newtxtext,newtxmath}
\usepackage[T1]{fontenc}

\DeclareRobustCommand{\VAN}[3]{#2}
\let\VANthebibliography\thebibliography
\def\thebibliography{\DeclareRobustCommand{\VAN}[3]{##3}\VANthebibliography}

\usepackage{graphicx}	% Including figure files
\usepackage{amsmath}	% Advanced maths commands
\title[Clumping in Virgo]{Gas clumping in the outskirts of the Virgo cluster}

\author[M. S. Mirakhor and S. A. Walker]{
M. S. Mirakhor\thanks{E-mail: msm0033@uah.edu}
and S. A. Walker
\\
Department of Physics and Astronomy, The University of Alabama in Huntsville, Huntsville, AL 35899, USA
}

\date{Accepted XXX. Received YYY; in original form ZZZ}

\pubyear{2015}

\begin{document}
\label{firstpage}
\pagerange{\pageref{firstpage}--\pageref{lastpage}}
\maketitle

\begin{abstract}
Observations of the ICM in the outskirts of the Virgo cluster with \textit{Suzaku} have found the gas mass fraction in the northern direction to be significantly above the expected level, indicating that there may be a very high level of gas clumping on small scales in this direction. Here we explore the \textit{XMM--Newton} data in the outskirts of Virgo, dividing it into a Voronoi tessellation to separate the bulk ICM component from the clumped ICM component. As the nearest galaxy cluster, Virgo’s large angular extent allows the spatial scale of the tessellation to be much smaller than has been achieved using the same technique on intermediate redshift clusters, allowing us to probe gas clumping on the scales down to 5$\times$5 kpc. We find that the level of gas clumping in the outskirts to the north is relatively mild, ($\sqrt{C} < 1.1$), suggesting that our point-source detection procedure may have excluded a significant fraction of clumps. While correcting for clumping brings the gas mass fraction at $r_{200}$ into agreement with the universal gas mass fraction, the values outside $r_{200}$ remain significantly above it. This may suggest that non-thermal pressure support in the outskirts to the north is significant, and we find that a non-thermal pressure support at the level of 20 per cent of the total pressure outside $r_{200}$ can explain the high gas mass fraction to the north, which is in agreement with the range expected from simulations.

%Measurements of the outskirts of the Virgo cluster have found that the gas mass fraction in the northern parts of the cluster is significantly higher than the universal baryon fraction, suggesting that there is a high level of gas clumping in the northern outskirts. To explore the  use a large mosaic of \textit{XMM--Newton} observations, we investigate gas clumping down to scales of $5 \times 5$ kpc for a representative region in the outskirts of the Virgo cluster, providing the first high-resolution view of gas clumping. When masking the point sources and the prominent substructures in the cluster field, we find that the recovered clumping factor profile along the northern strip is very mild ($\sqrt{C} < 1.1$), and show a slight trend of increasing with radius, implying that the bias introduced by clumping is low. We also find that the clumping factor measurements are highly uniform in all directions, and are not enhanced along the north-south direction. At $r_{200}$, the clumping-corrected gas mass fraction along the northern strip is statistically consistent with the universal gas mass fraction but, in the radial range of about (1.0--1.2)$r_{200}$, our gas mass fraction measurement is still significantly higher than the universal value. In this radial range, it is found that a support of about 20 per cent from the non-thermal to total pressure is required to reproduce the expected value of the universal value in the outskirts of the northern strip of the Virgo cluster, agreeing with the predictions of numerical simulations.

\end{abstract}


%%%%%%%%%%%%%%%%%%%%%%%%%%%%%%%%%%%%%%%%%%%%%%%%%%

%%%%%%%%%%%%%%%%% BODY OF PAPER %%%%%%%%%%%%%%%%%%

\section{Introduction}
Over the last decade, our broad-brush understanding of the physics of the intracluster medium (ICM) in galaxy cluster outskirts has improved significantly \citep[e.g.][]{Walker2013_Centaurus,simionescu2013thermodynamics,Urban2014,Tchernin2016,Ghirardini2019}. For a detailed review, see \citet{Walker2019}. One key factor is the role played by gas clumping, which is expected from simulations \citep[e.g.][]{Nagai2011,vazza2013properties,roncarelli2013large,planelles2017pressure,angelinelli2021properties} to increase in importance in the outskirts. These bright clumps can bias measurements of the gas density and, hence, the gas mass fraction high if they are not taken into account. However, the majority of our understanding is limited to azimuthally averaged studies, and on relatively large scales. The precise microphysics underlying gas clumping remains unexplored observationally, and can only be unravelled through observations of the nearest, brightest galaxy clusters, which afford us the highest spatial resolution view of clusters.

\begin{figure}
	\includegraphics[width=\columnwidth]{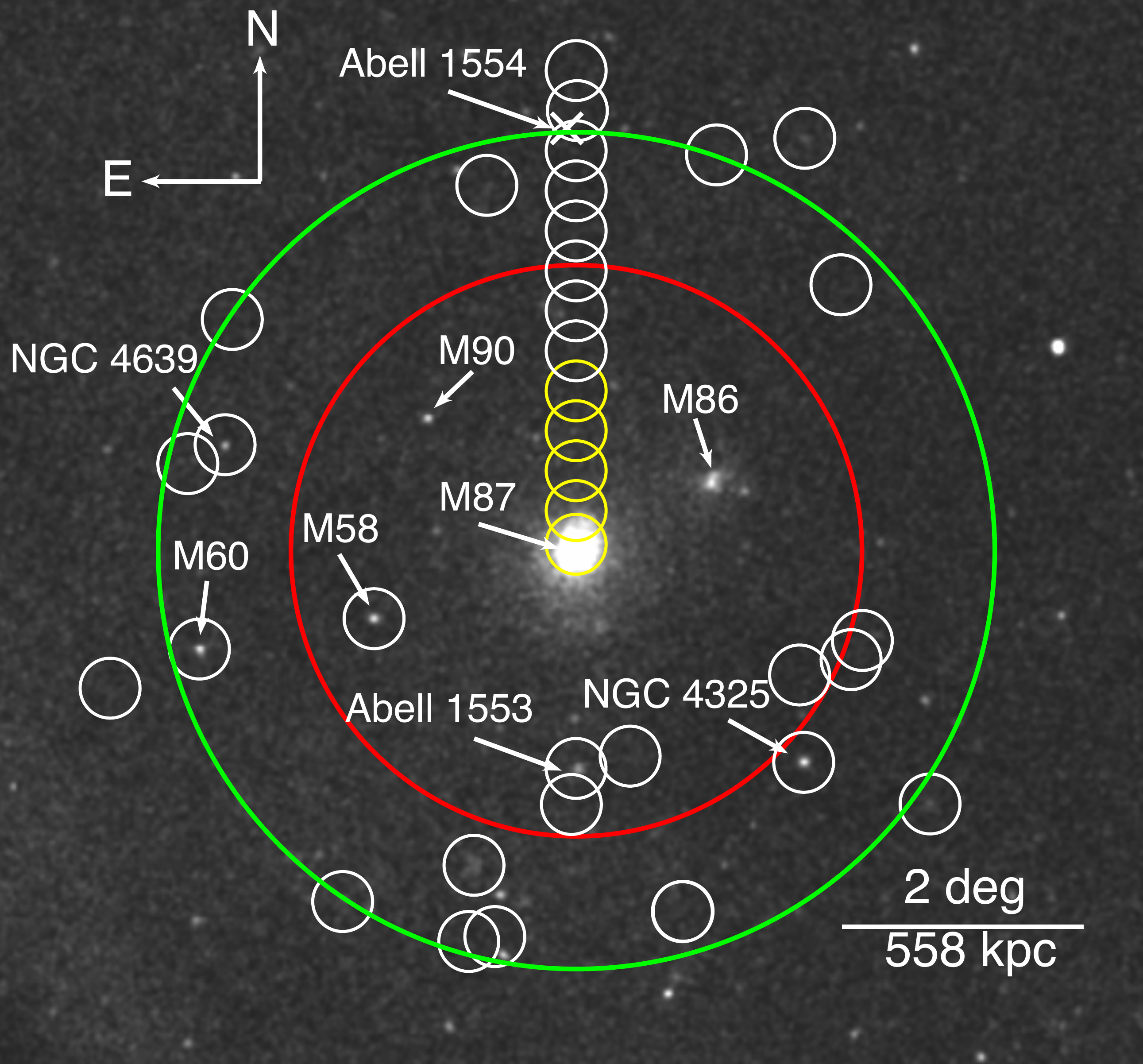}
	\caption{Adaptively smoothed \textit{ROSAT} All-Sky Survey image of the Virgo cluster in the 0.5$-$2.0 keV energy band with the positions of the \textit{XMM--Newton} observations studied in the current work being overlaid as white circles. The locations of the 13 \textit{XMM} pointings studied by \citet{Urban2011} covering the Virgo cluster from its centre northwards out to the virial radius are shown as yellow and white circles. The prominent substructures in the Virgo field are shown in the image. The red and green circles mark, respectively, the locations of the $r_{500}$ and $r_{200}$ radii. }
	\label{fig: Virgo_rosat}
\end{figure}

At a distance of only 16.1 Mpc, the Virgo cluster is the nearest galaxy cluster to us. Its large angular extent, with 1 arcmin corresponding to 4.65 kpc, provides a unique opportunity to study the behaviour of the ICM on scales far smaller than is achievable in any other galaxy cluster. In addition, the low temperature of the ICM in the outskirts of Virgo ($\sim$ 1 keV), combined with Virgo’s high X-ray flux, makes the Fe L line strong enough that the gas temperature, metal abundance and density can be obtained through spectral fitting out to $r_{200}$. Virgo is the only galaxy cluster for which these measurements can be made in the outskirts using \textit{XMM--Newton}. \citet{Urban2011} exploited this to analyze a strip of observations to the north, as shown in Fig. \ref{fig: Virgo_rosat}.

Virgo’s proximity allows the phenomenon of gas clumping to be probed on the smallest possible scale of all galaxy clusters. The level of gas clumping, and the spatial scale it operates on, remains uncertain observationally as only the very brightest gas clumps are bright enough to be resolved directly in X-rays. In simulations, the level of gas clumping is very sensitive to the level of gas cooling and the active galactic nucleus (AGN) feedback used in the simulations, with gas cooling acting to remove gas from the X-ray emitting phase, and AGN feedback acting to spread out gas and reduce the level of the simulated gas clumping  \citep[e.g.][]{vazza2013properties}.

It is possible to quantify the level of gas clumping \citep{eckert2015gas} by Voronoi tessellating the X-ray image, and plotting a histogram of the surface brightness values of each part of the tessellation. Clumping leads to a bright tail in the histogram, which systematically biases the mean surface brightness high, but has little effect on the median surface brightness in a given region. This is supported by simulations, which indicate that the ICM can be separated into a bulk component and a bright high-density tail from gas clumping \citep{zhuravleva2013quantifying}. The level of clumping can therefore be quantified by comparing how much larger the mean surface brightness is to the median surface brightness.

One limitation of the tessellation approach to exploring gas clumping is how small the tessellated regions can be made. Each region of the tessellation needs to contain at least 20 counts to allow for sufficiently accurate measurement of the surface brightness in that region. Since the gas clumps cannot be observed directly, it is at present unknown on what spatial scale this gas clumping operates on. If the clumping operates on scales smaller than the tessellation regions, the full magnitude of the clumping will be washed out and underestimated.

Most clusters studied so far in the outskirts with \textit{XMM--Newton} \citep[e.g.][]{ghirardini2018xmm,Ghirardini2019} have intermediate redshifts around 0.06, for which the typical size of a tessellated region containing 20 counts in the outskirts is around 1 arcmin$^{2}$, corresponding to a physical scale of around 65 $\times$ 65 kpc. For the Virgo cluster, this $65 \times 65$ kpc region covers an area of $14 \times 14$ arcmin. This provides us with a view of the surface brightness substructure that is over 100 times better resolved than intermediate redshift clusters. The typical size of a tessellated region containing 20 counts in the Virgo outskirts is around $1 \times 1$ arcmin, corresponding to a physical scale of around $5 \times 5$ kpc, allowing us to probe the effects of gas clumping on very fine scales. These detailed observations of the small scale surface brightness distribution are vital for improving our understanding of gas clumping, and for constraining numerical simulations. At present, we can only compare with simulations on spatial scales of around 60 kpc due to the intermediate redshift of the clusters that have been studied.

Using the \textit{Suzaku} data in 4 thin strips from the centre to the outskirts, \citet{Simionescu17} found that the gas density appears to be systematically higher in the outskirts to the north and south. Beyond $r_{200}$, the inferred gas mass fraction by \citet{Simionescu17} along the northern strip exceeds significantly the universal gas mass fraction, suggesting enhanced gas density possibly due to a filament candidate running in the north-south direction, connecting the cluster to the surrounding cosmic web. The accretion into galaxy clusters is expected to be highly asymmetrical, with different behaviour in the outskirts depending on whether accretion is proceeding along large scale filaments or from the low density gas from the surrounding void regions. Numerical simulations \citep[e.g.][]{zinger2016role} show that gas accreting from filamentary gas streams can penetrate deep into cluster interiors, carrying low entropy gas from the cosmic web deep inside them. Simulations \citep[e.g.][]{vazza2013properties} also predict that the level of gas clumping should be higher in these filament directions, however, this is yet to be confirmed observationally. 

In this work, we aim to probe gas clumping down to scales of $5 \times 5$ kpc in the outskirts of the Virgo cluster, providing the first high-resolution view of gas clumping, to see whether the high gas mass fraction measured in the northern strip in \citet{Simionescu17} can be explained by high levels of gas clumping on these scales. Using multiple observations in the region, we investigate the azimuthal variation in gas clumping around the virial radius of Virgo to test whether clumping is enhanced along the north-south direction, compared to other directions. Finally, we quantify the bias introduced by gas clumping on measurements of the gas mass fraction along the northern strip, allowing us to disentangle the effect of clumping from non-gravitational processes on very fine scales.

Throughout this work, we adopt a $\Lambda$ cold dark matter cosmology with $\Omega_{\rm{m}}=0.3$, $\Omega_{\rm{\Lambda}}=0.7$, and $H_0=100\,h_{100}$ km s$^{-1}$ Mpc$^{-1}$ with $h_{100}=0.7$. At the redshift of Virgo, 1 arcmin corresponds to 4.65 kpc. Uncertainties are at the 68 per cent confidence level, unless otherwise stated. For consistency with previous studies, we consider $r_{200}$, the radius within which the mean density is 200 times the critical density of the Universe, as the virial radius.

%----------------------
%----------------------

\section{data and analysis procedure}
\label{sec: data reduction}
In this paper, a total of 33 archival \textit{XMM--Newton} observations in the field of the Virgo cluster were used. 32 pointings cover mainly the outskirts of Virgo between 0.75$r_{500}$ and $r_{200}$, whereas 1 pointing lies far beyond the virial radius of the cluster, and was used to measure the local sky background. Table \ref{table: data} provides a list of all \textit{XMM--Newton} observations used in this work. In Fig. \ref{fig: Virgo_rosat}, we show the positions of the \textit{XMM--Newton} observations as white circles overlaid on the adaptively smoothed \textit{ROSAT} All-Sky Survey image of the Virgo cluster in the soft energy band. The X-ray data were reduced using \textit{XMM--Newton} Science Analysis System (\textit{XMM}-SAS) version 18.0 and current calibration files (CCF), following the method described in the Extended Source Analysis Software (ESAS) cookbook\footnote{https://heasarc.gsfc.nasa.gov/docs/xmm/esas/cookbook/xmm-esas.html}, as is also done in \citet{mirakhor2020complete,mirakhor2020high} and \citet{mirakhor2021exploring}. Below, we provide an outline of the data reduction procedure performed on the \textit{XMM--Newton} observations.

%----------------------------
\begin{table*}
\begin{minipage}{130mm}
    \centering
    \caption{\textit{XMM--Newton} observations of Virgo}
    \begin{tabular}{lccccc}
    \hline
    Obs. ID   & RA     & Dec.    & Obs. Date  & Exposure (ks)  & Offset\footnote{Distance from the centre of Virgo, which is centred on the galaxy M87 (RA = 12:30:49.4, Dec. = +12:23:28.0).} (arcmin)      \\
        \hline
0106060501   & 12 30 50.00 & +14 03 30.0 & 2002-07-06  & 17.74  & 100.0 \\
0414980101   & 12 28 59.91 & +10 40 32.9 & 2007-06-10  & 23.92  & 106.4 \\
0827011101   & 12 30 50.10 & +10 34 27.8 & 2018-06-10  & 36.00  & 109.0 \\
0106060601   & 12 30 50.00 & +14 23 30.0 & 2002-07-08  & 14.88  & 120.0 \\
0556210601   & 12 30 59.72 & +10 16 24.8 & 2008-06-13  & 8.92   & 127.1 \\
0651790301   & 12 23 13.74 & +11 20 52.1 & 2010-06-02  & 31.42  & 127.9 \\
0106060701   & 12 30 50.00 & +14 43 30.0 & 2002-07-05  & 14.38  & 140.0 \\
0802580101   & 12 21 28.20 & +11 28 18.6 & 2017-12-04  & 34.00  & 147.9 \\
0782010201   & 12 21 04.99 & +11 37 52.4 & 2016-06-10  & 46.00  & 150.0 \\
0108860101   & 12 23 06.68 & +10 37 10.0 & 2000-12-24  & 22.11  & 155.4 \\
0106061401   & 12 30 50.00 & +15 03 30.0 & 2002-06-13  & 8.88   & 160.0 \\
0673851101   & 12 34 17.43 & +09 45 58.1 & 2012-01-02  & 39.91  & 165.6 \\
0106060901   & 12 30 50.00 & +15 23 30.0 & 2002-06-10  & 15.62  & 180.0 \\
0112551001   & 12 42 51.98 & +13 15 26.0 & 2001-12-16  & 14.96  & 183.6 \\
0306060101   & 12 21 42.51 & +14 35 52.0 & 2005-12-05  & 96.45  & 187.6 \\
0651790101   & 12 27 13.24 & +09 22 59.4 & 2010-06-13  & 28.61  & 188.1 \\
0404120101   & 12 33 55.47 & +15 26 07.8 & 2006-12-19  & 31.91  & 188.2 \\
0802580201   & 12 27 14.51 & +09 22 48.8 & 2017-12-05  & 34.14  & 188.2 \\
0502160101   & 12 43 39.61 & +11 33 09.0 & 2007-12-19  & 91.97  & 195.0 \\
0802580401   & 12 33 36.55 & +09 10 59.4 & 2017-07-15  & 36.00  & 196.8 \\
0651790201   & 12 44 07.87 & +13 05 45.7 & 2010-06-13  & 28.92  & 199.2 \\
0106061001   & 12 30 50.00 & +15 43 30.0 & 2002-06-06  & 15.72  & 200.0 \\
0722960101   & 12 34 27.52 & +09 07 54.2 & 2013-06-30  & 54.00  & 202.8 \\
0202730301   & 12 42 40.09 & +14 18 07.7 & 2004-06-13  & 38.91  & 207.4 \\
0504240101   & 12 42 40.09 & +14 18 07.7 & 2007-06-06  & 93.82  & 207.4 \\
0802580301   & 12 25 58.36 & +15 41 12.0 & 2017-06-23  & 35.50  & 210.0 \\
0504100601   & 12 38 43.40 & +09 27 37.0 & 2007-12-09  & 21.92  & 210.9 \\
0843831001   & 12 18 50.50 & +10 15 54.0 & 2019-06-16  & 25.70  & 217.6 \\
0106061101   & 12 30 50.00 & +16 03 30.0 & 2002-06-09  & 16.62  & 220.0 \\
0106860201   & 12 22 54.98 & +15 49 12.0 & 2001-12-28  & 36.63  & 235.7 \\
0106061201   & 12 30 50.00 & +16 23 30.0 & 2002-06-09  & 18.44  & 240.0 \\
0803952801   & 12 46 40.36 & +11 13 02.9 & 2017-07-03  & 40.40  & 243.1 \\
0106061301\footnote{Local background pointing.}   & 12 30 50.00 & +17 53 30.0 & 2002-06-10  & 16.62  & 330.0 \\
  \hline
    \end{tabular}
    \vspace{-5mm}
    \label{table: data}
\end{minipage}
\end{table*}

\subsection{Data reduction}
\label{sec: data_reduction}
We processed the data by running the \textit{epchain} and \textit{emchain} tasks, followed by the \textit{mos-filter} and \textit{pn-filter} tasks to remove soft proton flares and create clean event files. We examined the MOS detectors for CCDs in anomalous data, and affected CCDs were then excluded from the analysis. We then created spectra, response files, and exposure maps by running the \textit{mos-spectra} and \textit{pn-spectra} scripts. From these intermediate files, we then created the quiescent particle background spectra and images in the MOS and PN coordinates by running the \textit{mos-back} and \textit{pn-back} tasks. The data were further examined for residual soft-proton contamination by running the ESAS task \textit{proton}.

\begin{figure}
    \centering
	\includegraphics[width=0.7\columnwidth]{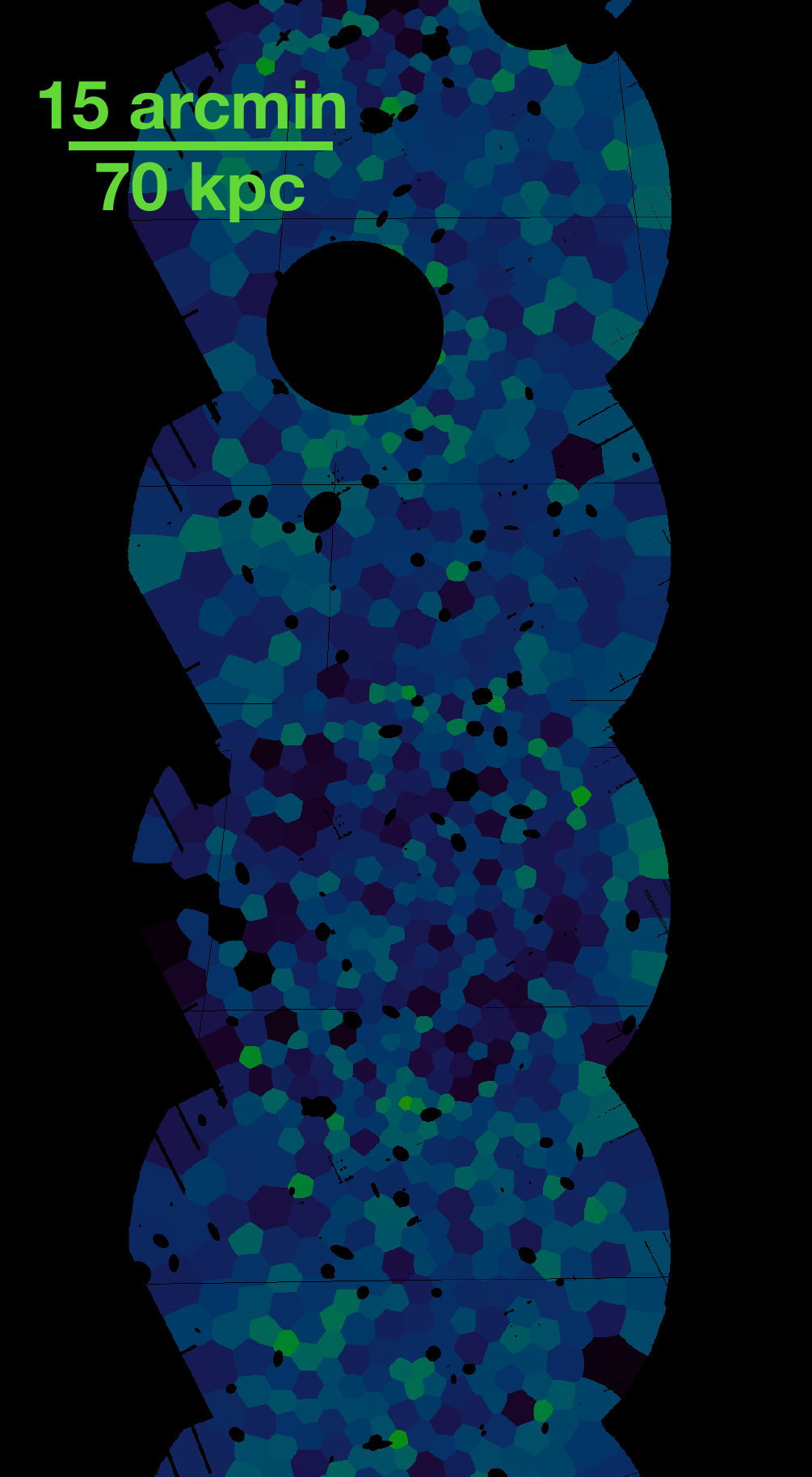}
	\caption{Voronoi tessellated image of 4 pointings along the northern strip of the Virgo cluster in the 0.7$-$1.2 keV energy range, covering the outskirts of the cluster between $r_{500}$ and $r_{200}$. The typical area of a tessellated region in the outskirts is around $1 \times 1$ arcmin, which corresponds to a physical area of around $5 \times 5$ kpc at the redshift of the Virgo cluster. This allows us to explore the effects of gas clumping on very fine scales, of two orders of magnitude finer than can be achieved by exploring intermediate redshift clusters.}
	\label{fig: north_pointings}
\end{figure}

We carried out the analysis procedure described above for each \textit{XMM--Newton} observation listed in Table \ref{table: data}. After weighting the \textit{XMM-Newton} instruments by their effective area, the main components required to create a background-subtracted and exposure-corrected image were stacked and adaptively smoothed into a single image. In Fig. \ref{fig: north_pointings}, we show a background-subtracted and exposure-corrected image of 4 pointings along the northern strip of Virgo in the 0.7$-$1.2 keV energy range, covering the outskirts of the cluster between $r_{500}$ and $r_{200}$. We chose this energy range to maximize the signal-to-noise ratio in the outskirts. A Vorronoi tessellation algorithm \citep{diehl2006adaptive} was applied on the mosaicked count image to create an adaptively binned image with each tessellated region contains at least 20 counts, allowing for accurate measurement of the X-ray surface brightness in the outskirts. The typical area of a tessellated region in that region is around $1 \times 1$ arcmin, which corresponds to a physical area of around $5 \times 5$ kpc at the redshift of the Virgo cluster. This provides us with a view of the surface brightness substructure that is over 100 times better resolved than can be achieved by exploring intermediate redshift clusters.

%----------------------
%----------------------

\subsection{Point source detection}
\label{sec: point_source}
We initially run the ESAS source-detection tool \textit{cheese} to detect point sources and extended substructures down to $5\sigma$ significance that contaminated the field of view. We then run the \textit{Chandra} tool \textit{wavdetect} with  wavelet scales of 2, 4, 8, 16, 32, and 64 pixels to detect remaining point sources within the field that were missed using the \textit{cheese} task. Furthermore, we visually inspected the mosaicked image of Virgo to correct for false detections, and point sources that were missed by \textit{cheese} and \textit{wavdetect}. A total of 1403 point sources and extended substructures were detected in the field. These sources were then excluded from the subsequent analysis. 

We note that only a small fraction of the detected point sources are spectroscopically confirmed members of the Virgo cluster, while others are projected into the cluster extended X-ray emission. The typical radius of the vast majority of the detected point sources is in the range of 10–30 arcsec. However, we used a larger extraction radius to exclude emission from some of the bright point sources and extended substructures. Most of these sources, as labelled in Fig. \ref{fig: Virgo_rosat}, are galaxies and galaxy clusters, residing outside the Virgo cluster at higher redshifts.

\section{Method}
\label{sec: method}
\citet{zhuravleva2013quantifying} showed that the gas density distribution in a given region of the ICM follows approximately a log-normal distribution, and a high-density tail resulting from denser outliers or gas clumping. These authors found that the median of the density distribution coincides well with the peak of the log-normal component, and hence it is robust against outliers. \citet{eckert2015gas} then suggested to use the median of the surface brightness to recover the true surface brightness in the presence of gas clumping. Therefore, as the mean of the surface brightness tends to bias the true surface brightness, we can estimate the clumping factor directly from the X-ray image by dividing the mean of the deprojected X-ray surface brightness profile by the median of the same profile in a given annulus, as is done in \citet{eckert2015gas}. 

In order to estimate the clumping factor, we need to resolve the shape of the surface brightness distribution into the log-normal distribution and the tail in each region we study. To do that, a sufficient number of Voronoi tessellated regions is required. We find that a minimum of 150 independent Voronoi regions is sufficient to be able to accurately calculate the mean and median of the surface brightness distribution, and therefore the clumping factor. We ensure that this threshold is met for all of the regions we study in the current work.   

Due to Virgo’s close proximity, the effect of gas clumping can be explored on very fine scales. We can resolve clumps on scales larger than the half energy width of the point spread function (PSF; 17 arcsec for MOS, corresponding to 1.3 kpc at the redshift of Virgo). This allows to study the properties of the ICM on scales far smaller than is achievable in any other galaxy cluster, and thus to improve the precision of the X-ray measurements of the ICM.

\section{Results}
\label{sec: results}
\subsection{Northern strip}
\label{sec: northern_strip}
\begin{figure}
	\includegraphics[width=\columnwidth]{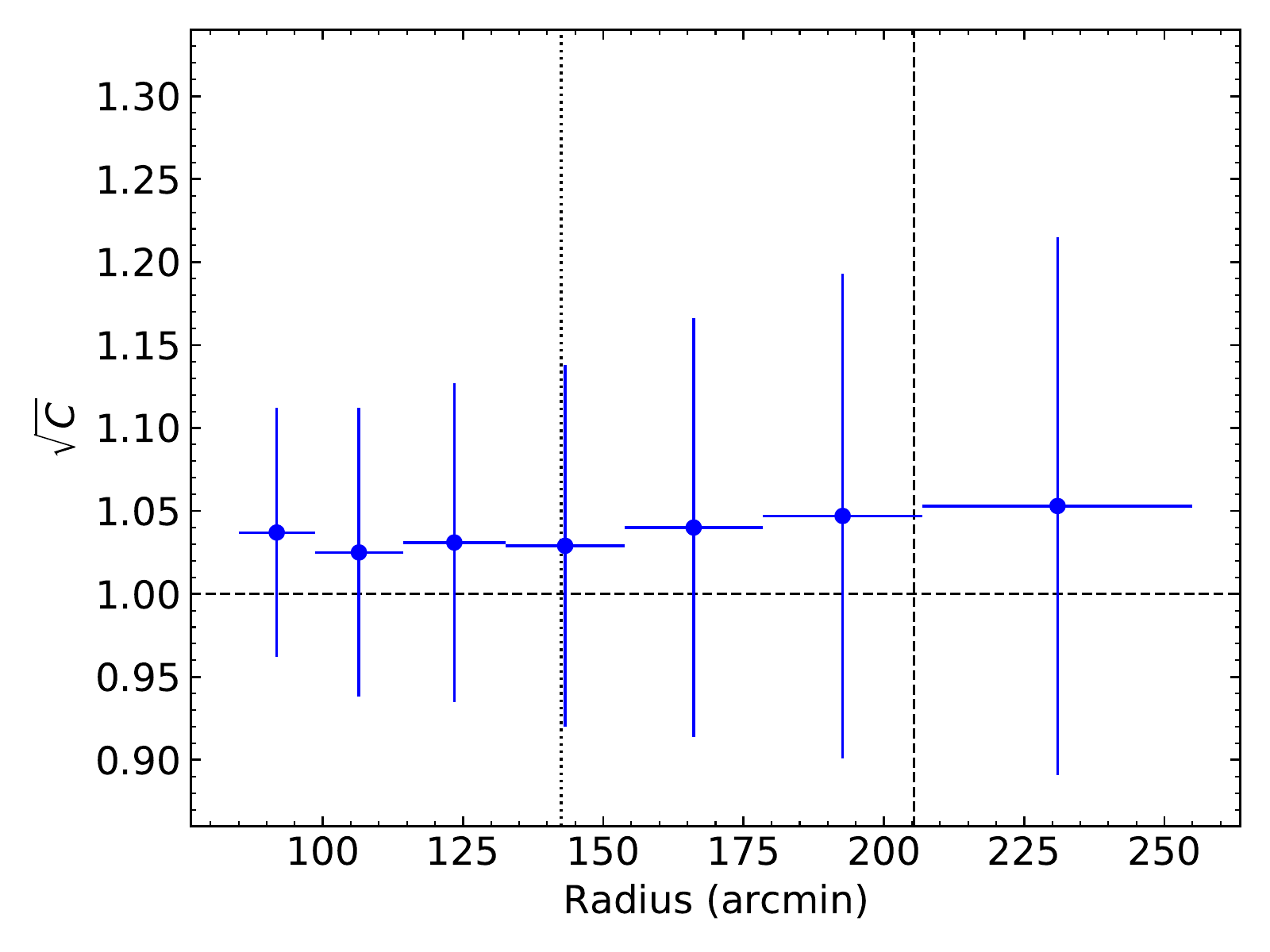}
	\caption{Radial profile of the clumping factor along the northern strip of the Virgo cluster. The vertical dotted and dashed lines represent the $r_{500}$ and $r_{200}$ radii, respectively. The horizontal dashed line marks the value of the clumping factor for a perfectly uniform ICM. Our results indicate that the values of the recovered clumping factor are low out to the cluster outskirts, and show a slight trend of increasing with radius. }
	\label{fig: clumping_N}
\end{figure}

\begin{figure}
	\includegraphics[width=\columnwidth]{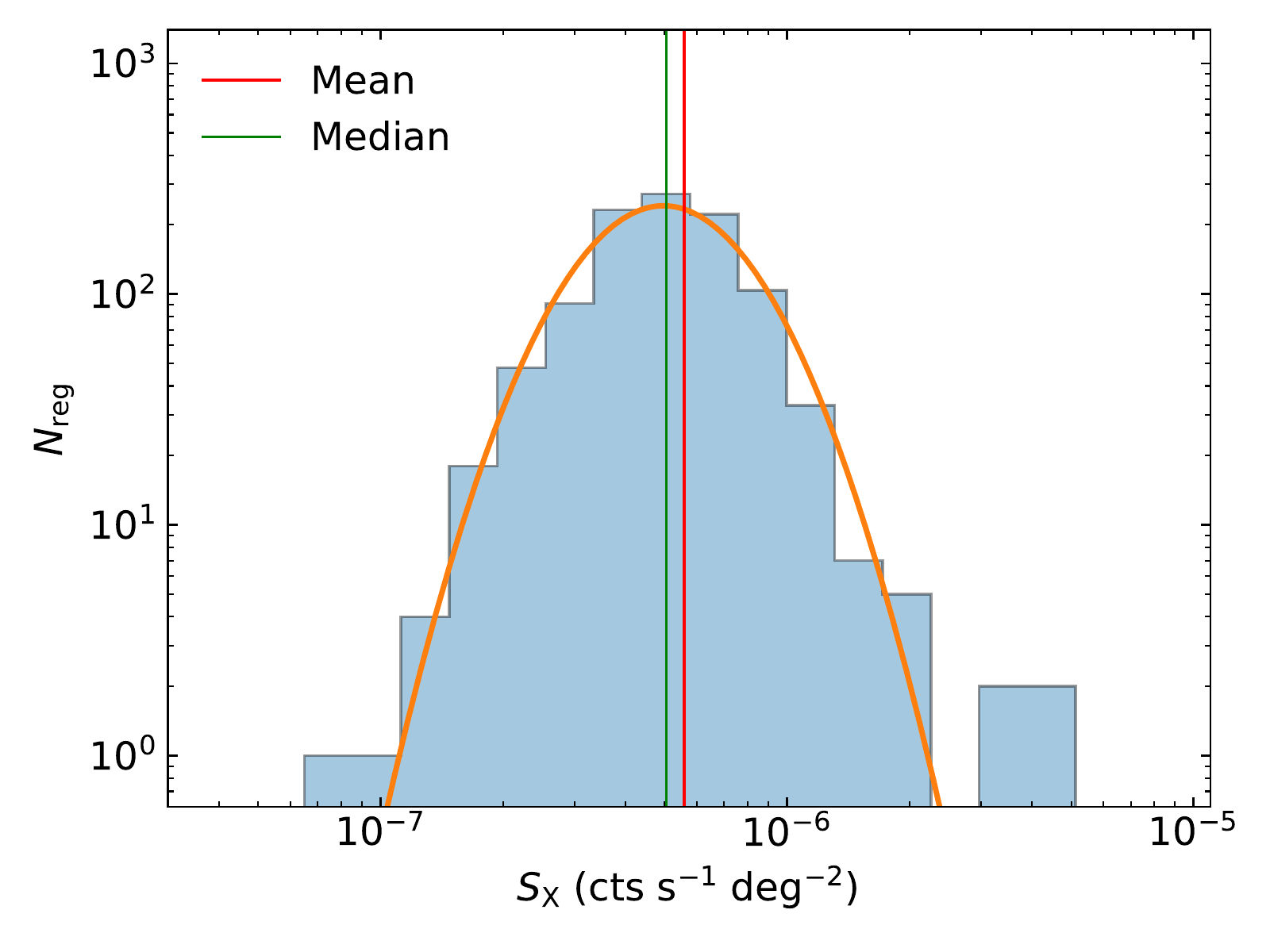}
	\caption{Distribution of the X-ray surface brightness on scales of $5\times5$ kpc for the region along the northern strip of the Virgo cluster, showing that the surface brightness distribution of the ICM can be divided into a bulk component and a high-density tail resulting from gas clumping. The orange curve is the best-fitting log-normal distribution. The median surface brightness (green line) coincides better than the mean (red line) with the peak of the best-fitting model.   }
	\label{fig: histogram_N}
\end{figure}

Fig. \ref{fig: clumping_N} shows the estimated values of the clumping factor along the northern strip of the Virgo cluster on scales of 5$\times$5 kpc. The error on the clumping factor is estimated through Monte Carlo simulations by generating $10^4$ random realizations. The $1 \sigma$ confidence interval is computed from the distribution of the different realizations. We find that the values of the recovered clumping factor are low ($\sqrt{C} < 1.1$) out to the cluster outskirts, and show a slight trend of increasing with radius. This implies that the clumping bias is very mild in the northern strip of the Virgo cluster on scales of 5$\times$5 kpc. In Fig. \ref{fig: histogram_N}, we show a histogram of the distribution of the surface brightness on scales as small as $5 \times 5$ kpc for the region along the northern strip of the Virgo cluster. The surface brightness distribution is fitted with a log-normal distribution. The figure shows that the observed distribution can be broadly divided into a bulk component and a high-density tail resulting from gas clumping, as predicted by numerical simulations \citep[e.g.][]{zhuravleva2013quantifying}. It is clear that only few regions have relatively high surface brightness, implying that gas clumping is low, and therefore the clumping bias is very mild in the northern strip on the small scales that we study.

%----------------------
%----------------------

%----------------------
%----------------------

\subsection{Azimuthal variation in clumping}
\label{sec: clumping_variation}
Using \textit{Suzaku} data, \citet{Simionescu17} presented the detailed spectroscopic measurements of the thermodynamical properties in 4 thin strips from the centre to the outskirts of the Virgo cluster. They found that the gas density appears to be systematically higher along the north–south direction, implying that there is a large-scale structure filament running in this direction, connecting the cluster to the surrounding cosmic web.

To test whether the observed clumping is enhanced along the north-south direction compared to the rest of the ICM, it requires to have a knowledge of gas clumping at multiple sectors in the outskirts of Virgo to measure the azimuthal variation of the clumping factor. At present, the area of the Virgo cluster covered by \textit{XMM--Newton} in the outskirts is low, as shown in Fig. \ref{fig: Virgo_rosat}, where we show the existing pointings as the white circles. However, it is still possible to measure the clumping factor at multiple sectors if we ensure that enough independent Voronoi tessellated regions are available in each azimuthal sector we study. 

\begin{figure}
	\includegraphics[width=\columnwidth]{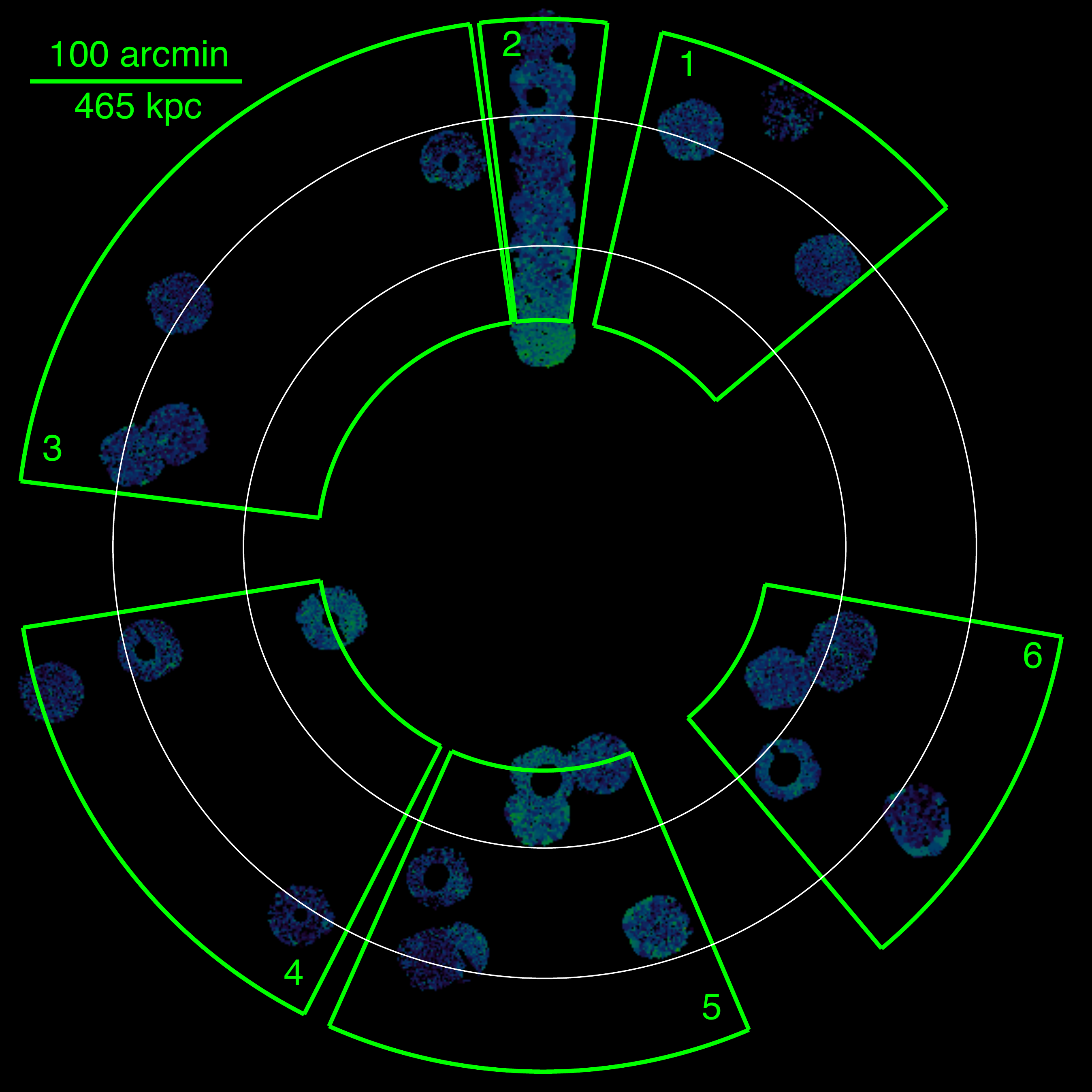}
	\caption{Locations of the 6 azimuthal sectors overlaid on the \textit{XMM} image of the Virgo cluster, covering the radial range between 0.75$r_{500}$ and 1.20$r_{200}$. Every sector has at least 500 independent Voronoi regions in order to accurately measure the shape of the surface brightness, and therefore the clumping factor. The inner and outer white circles mark the locations of $r_{500}$ and $r_{200}$, respectively. }
	\label{fig: sectors}
\end{figure}

\begin{figure}
	\includegraphics[width=\columnwidth]{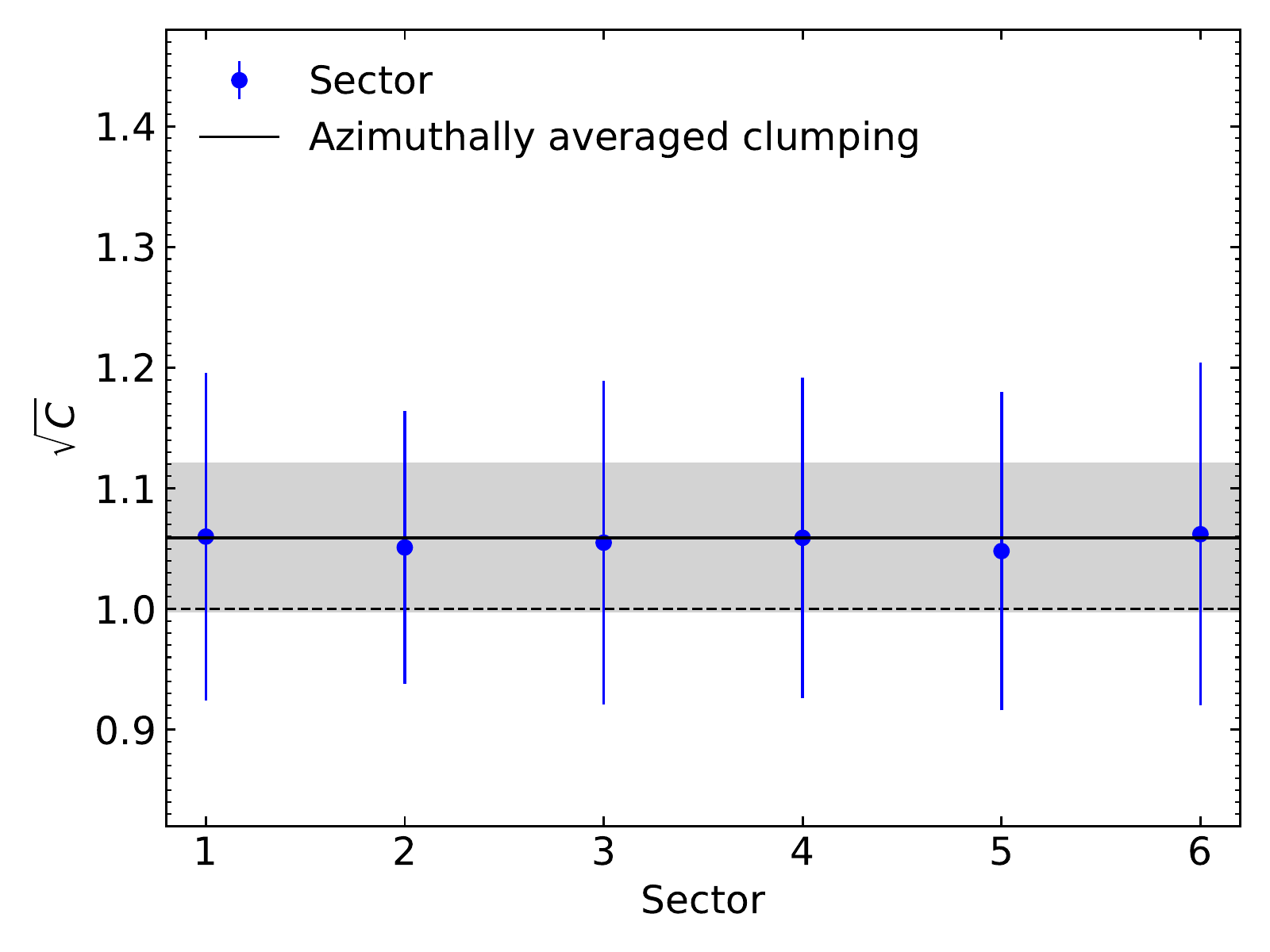}
	\caption{Azimuthal variation in the gas clumping in the radial range between 0.75$r_{500}$ and 1.20$r_{200}$ for all of the azimuthal sectors. Since the \textit{XMM-Newton} coverage in the outskirts is low, the recovered clumping factors are not representing the entire sector, but only the regions where the X-ray coverage is available. It is clear that the clumping factor measurements are highly uniform in all directions, and are not enhanced along the north-south direction.}
	\label{fig: Clumping_sectors}
\end{figure}

We, therefore, divided the outskirts of the Virgo cluster into 6 sectors such that every sector has at least 500 independent Voronoi regions. All galaxies and resolved point sources have been removed from the mosaic. In Fig. \ref{fig: sectors}, we show the locations of the azimuthal sectors overlaid on the \textit{XMM} image of the Virgo cluster, covering the radial range between 0.75$r_{500}$ and 1.20$r_{200}$. Fig. \ref{fig: Clumping_sectors} shows the azimuthal variation in the clumping factor for all of the azimuthal sectors down to scales of 5$\times$5 kpc. In the same figure, we also show the azimuthally averaged clumping factor in the same radial range. Since the \textit{XMM--Newton} coverage in the outskirts is low, the recovered clumping factors are not representing the entire sector, but only the regions where the X-ray coverage is available. This figure clearly shows that the clumping factor measurements are highly uniform in all directions, and are not enhanced along the north-south direction. 

Alternatively, it is still possible that gas clumping is enhanced along the north-south direction, possibly due to the existence of a large-scale filament, but our point-source detection procedure may have been able to detect most of the point sources and the prominent substructures in the field of the cluster down to scales of $5 \times 5$ kpc in all directions. This would explain the moderate level of gas clumping, and the good agreement among the measurements of the clumping factors seen along different directions in the outskirts of the Virgo cluster, since all of the clumps would have been resolved out by our point source detection algorithm (Fig. \ref{fig: Clumping_sectors}).

The estimated values of the clumping factor in the above analysis were determined by masking all point sources and extended substructures in the field of Virgo. However, we also determined the clumping factor by keeping the extended substructures in the field, i.e., masking only the point sources. By doing that, we find that the clumping factor ($\sqrt{C}$) values increase significantly, reaching to around 1.2--1.5 in the outskirts. Furthermore, we smeared the \textit{XMM} image of the Virgo cluster with the \textit{Suzaku} PSF. We find that doing this causes the clumping factor to rise to around 1.2 near the outskirts, showing that the \textit{Suzaku} PSF can cause an overestimate in the true value of gas clumping.

%----------------------
%----------------------

%----------------------
%----------------------

\subsection{Correcting the gas mass fraction bias}
\label{sec: correcting_fg_bias}
The gas mass fraction, the ratio of the gas to total mass, in a galaxy cluster provides a valuable probe for cosmological studies \citep[e.g.][]{Allen2008,ettori2009cluster}. The gas mass fraction of galaxy clusters is expected to reflect the universal baryon fraction \citep[e.g.][]{white1993baryon}. In many clusters, however, the recovered value for the gas mass fraction differs from the universal value \citep[e.g.][]{eckert2013x,ghirardini2018xmm}. This difference is mainly due to the presence of several systematics, including the presence of gas clumping. Therefore, the knowledge of gas clumping improves the gas mass fraction measurements of galaxy clusters. At the same time, it allows us to disentangle its effects from the non-thermal pressure component, which also bias measurements of the gas mass fraction.

%Gas clumping bias the gas density and, consequently, the gas and total mass measurements, the quantities from which the gas mass fraction is derived. In the presence of gas clumping, the X-ray gas mass can be biased high by a factor of $\sqrt{C}$, whereas the total mass can be biased low by the derivative of $\sqrt{C}$. Therefore, the bias in the gas mass fraction is expected to be relatively higher with respect to the gas and total mass quantities, as it includes the positive bias in the gas mass and the negative bias in the total mass. 

\begin{figure}
	\includegraphics[width=\columnwidth]{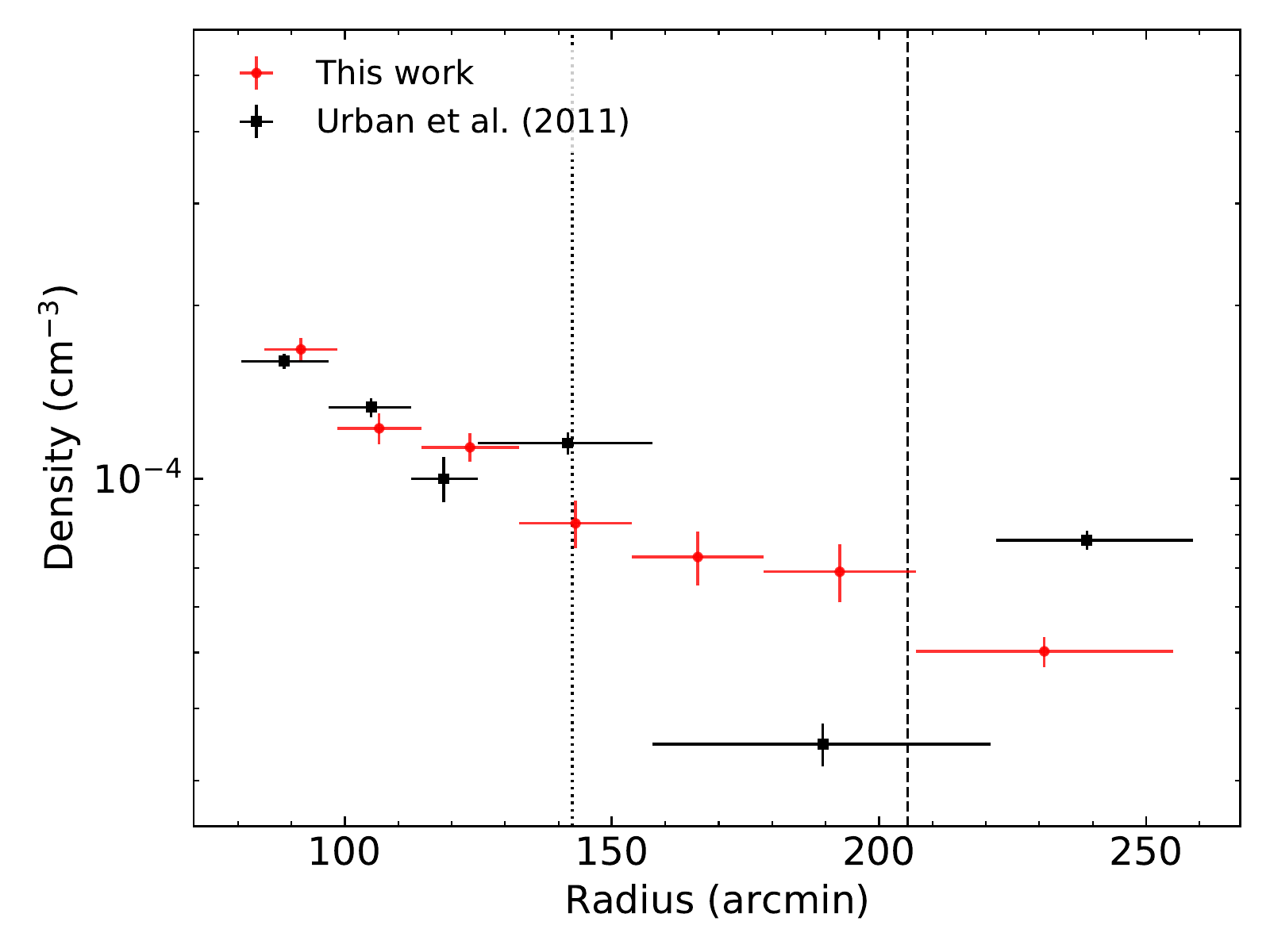}
	\caption{Clumping-corrected gas density (red circle points) along the northern strip of Virgo obtained by deprojection of the surface brightness under the assumption of spherical symmetry. The black square points show the gas density reported in \citet{Urban2011}. The vertical dotted and dashed lines represent the $r_{500}$ and $r_{200}$ radii, respectively.}
	\label{fig: density_N}
\end{figure}

To quantify the effect of gas clumping on the gas mass fraction measurements along the northern strip of Virgo, it requires to have an accurate knowledge of the gas density and temperature of the ICM. We measured the gas density along the northern strip using the \textit{XMM--Newton} data, following the same approach as \citet{Tchernin2016} and \citet{ghirardini2018xmm}, as is also done in \citet{mirakhor2020complete,mirakhor2020high} and \citet{mirakhor2021exploring}, where the deprojected X-ray surface brightness is converted into a deprojected gas density under the assumption of spherical symmetry, using the metal abundances measured in the outskirts in \citet{Urban2011}. In Fig. \ref{fig: density_N}, we show the radial profile of the clumping-corrected gas density along the northern strip of Virgo, compared to the gas density reported in \citet{Urban2011}. The deprojected density profile from \citet{Urban2011} shown in Fig. \ref{fig: density_N} has an unusual oscillating behavior, with the density in their outermost bin much higher than in their penultimate bin. This type of effect can occur when the density in the outermost annulus is overestimated (possibly by not taking into account gas clumping), as this causes an overestimate of the emission projected onto the annulus inside it. This can cause the deprojected density in the annulus inside the outermost annulus to be underestimated (for examples of this `ringing’ effect in deprojections see \citealt{Russell2008}). By comparison, our outermost density is lower because we have taken into account the gas clumping correction, and our deprojected profile has a much more physical non-oscillating shape. For temperature, we adopted the \citet{Urban2011} measurements, which were obtained using the same \textit{XMM--Newton} data used in this work. The density and temperature measurements are then used to estimate the Virgo masses and, consequently, the gas mass fraction.

%----------------------
%----------------------

The gas mass can be estimated by integrating the deprojected gas density, $\rho(r)$, over a given volume:
%----------------------
\begin{equation}
 M_{\rm{gas, obs}}=4\pi \int \rho(r)r^2dr,
 \label{eq: gas_mass}
\end{equation}
%----------------------
where the integration is computed using a four-point spline interpolation and the limit is taken out to some cut-off radius, as is done in \citet{mirakhor2020complete,mirakhor2020high}. Since gas clumping affects the gas density, the gas mass is also affected by clumping. Thus, equation (\ref{eq: gas_mass}) takes the form:
%----------------------
\begin{equation}
 M_{\rm{gas, true}}=4\pi \int \sqrt{C(r)} \rho(r)r^2dr,
 \label{eq: gas_mass_corr}
\end{equation}
and this leads to a bias in the gas mass that can be expressed as
%---------------------
\begin{equation}
 b(M_{\rm{gas}})=\frac{M_{\rm{gas, true}} - M_{\rm{gas, obs}}}{M_{\rm{gas, obs}}},
\label{eq: gas_mass_bias}
\end{equation}
%-------------------
which is always $> 0$ in the presence of gas clumping.

For the total mass, its distribution relates to the gas density and temperature of the ICM. Assuming that the ICM is in the hydrostatic equilibrium in the cluster's gravitational potential well, the total mass enclosed in radius $r$ is
%-------------------
\begin{equation}
    M_{\rm{HSE,obs}}= - \frac{k_{\rm{B}} T(r) r}{G \mu m_{\rm{p}}}\bigg(\frac{{\rm{d}} \log \rho(r)}{{\rm{d}} \log r} + \frac{{\rm{d}} \log T(r)}{{\rm{d}} \log r}  \bigg),
    \label{eq: tot_mass}
\end{equation}
%------------------
where $T(r)$ is the gas temperature, $k_{\rm{B}}$ is the Boltzmann constant, $G$ is the Newtonian gravitational constant, and $\mu$ is the mean molecular weight in units of the proton mass, $m_{\rm{p}}$. We used a three-point quadratic interpolation to compute the derivatives in equation (\ref{eq: tot_mass}).

Determining the cluster mass under the hypothesis of hydrostatic equilibrium (equation \ref{eq: tot_mass}) is also affected by gas clumping. In the presence of gas inhomogeneities in the ICM, the total mass takes the form:
%-------------------
\begin{equation}
    M_{\rm{HSE,true}}= - \frac{k_{\rm{B}} T(r) r}{G \mu m_{\rm{p}}}\bigg(\frac{{\rm{d}} \log \rho(r)}{{\rm{d}} \log r} + \frac{{\rm{d}} \log T(r)}{{\rm{d}} \log r} + \frac{1}{2} \frac{{\rm{d}} \log C(r)}{{\rm{d}} \log r}  \bigg).
    \label{eq: tot_mass_corr}
\end{equation}
%------------------
Compared to $\rho(r)$ and $T(r)$, the clumping factor measurements show a very mild dependence on radius, as we can see in Fig. \ref{fig: clumping_N}. This implies that the cluster total mass is less affected by the presence of gas clumping.

However, the effect of gas clumping on the gas mass fraction is expected to be relatively higher with respect to the gas and total mass, as it includes the positive bias in the gas mass and the negative bias in the total mass. The bias in the gas mass fraction is 
%-------------------
\begin{equation}
b(f_{\rm{gas}}) = \frac{f_{\rm{gas,true}} - f_{\rm{gas,obs}}}{f_{\rm{gas,obs}}}=\frac{b(M_{\rm{gas}}) - b(M_{\rm{HSE}})}{1 + b(M_{\rm{HSE}})}.
\label{eq: fgas_bias}
\end{equation}
%----------------------

%----------------------
\begin{figure}
	\includegraphics[width=\columnwidth]{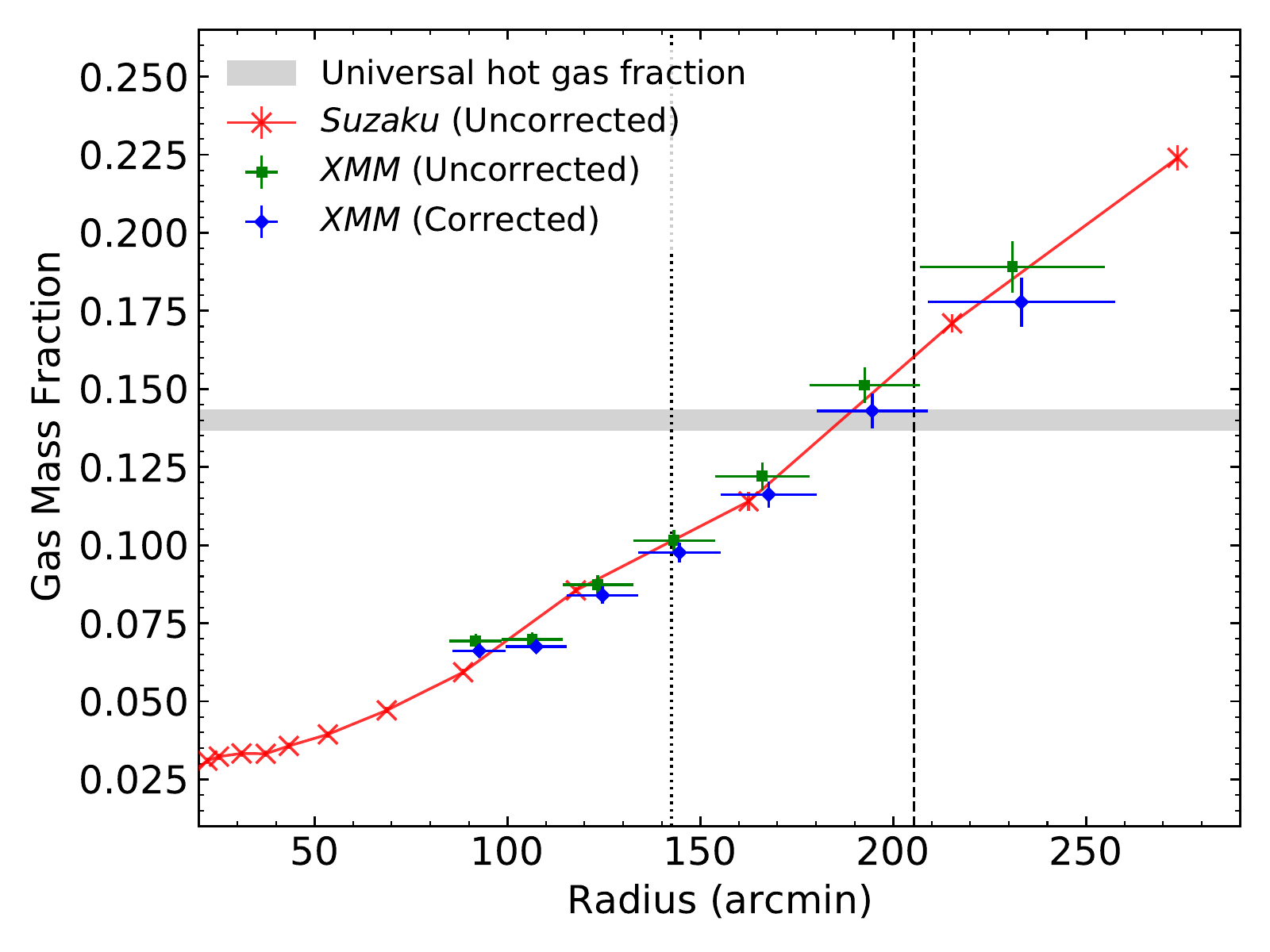}
	\caption{Radial profiles of the gas mass fraction along the northern strip of the Virgo cluster. The blue and green data points show the gas mass fraction measurements with and without the clumping correction. The red data points show the clumping-uncorrected gas mass fraction measurements obtained using \textit{Suzaku} \citep{Simionescu17}. The grey shadow area marks the universal hot gas mass fraction obtained from the cosmic baryon fraction \citep{ade2016planck}, and assuming that about 10 per cent of baryons are in stars. The vertical dotted and dashed lines represent the $r_{500}$ and $r_{200}$ radii, respectively. Even with the clumping correction, the gas mass fraction outside $r_{200}$ remains significantly above the universal hot gas mass fraction.  }
	\label{fig: fgas}
\end{figure}
%----------------------

Fig. \ref{fig: fgas} shows the radial profiles of the gas mass fraction along the northern strip of the Virgo cluster, with and without the clumping correction. In the same figure, we compare our measurements with the clumping-uncorrected gas mass fraction measurements along the northern strip obtained using \textit{Suzaku} \citep{Simionescu17}. This figure shows that the gas mass fraction measurements are systematically overestimated when the effect of gas clumping is neglected. This overestimate show a mild trend of increasing bias as we go further away from the cluster centre. In the outskirts of Virgo, we find that the gas mass fraction is overestimated by about 5 per cent, in agreement with the bias inferred by \citet{eckert2015gas} for a sample of galaxy clusters. %The estimated bias, however, is significantly lower than that reported for the north-west arm of the Perseus cluster \citep{Simionescu2011}.

In Fig. \ref{fig: fgas}, we also overplotted the value of the universal gas mass fraction estimated based on the cosmic baryon fraction of $0.156 \pm 0.003$ reported by \citet{ade2016planck}, and assuming that about 10 per cent of baryons are in stars \citep[e.g.][]{sanderson2013baryon,chiu2018baryon}. When accounting for the gas clumping bias, the gas mass fraction drops, agreeing with the universal hot gas mass fraction at $r_{200}$. However, our clumping-corrected gas mass fraction is significantly higher than the universal value beyond $r_{200}$ in the radial range of about 205--255 arcmin, implying that gas clumping is only partially responsible for the gas mass fraction excess seen in the region. If we adopt the $1 \sigma$ upper limit on the inferred clumping factor, then the gas mass fraction drops to $\sim 0.156$, which is still higher than the universal value. To reproduce the expected value of the universal gas mass fraction beyond $r_{200}$, we find that it requires the clumping factor $\sqrt{C} \approx 1.3$, corresponding roughly to the $2 \sigma$ upper limit on the recovered clumping factor.

Alternatively, the excess in the gas mass fraction measured in the radial range of about 205--255 arcmin might suggest that there is a substantial contribution from non-thermal pressure, which is expected to play an important role in the outskirts of galaxy clusters \citep[e.g.][]{nelson2014weighing,biffi2016nature}. This, in turn, might imply that the assumption of hydrostatic equilibrium is not valid in the outskirts of the northern strip. We, therefore, modify the hydrostatic equilibrium equation (equation \ref{eq: tot_mass}) by adding an extra component in order to account for non-thermal pressure. By setting $\alpha$ as the ratio of the non-thermal to total pressure and rearranging the terms in the equation, the total mass becomes
%----------------------
\begin{equation}
 M_{\rm{tot}}=\frac{1}{(1-\alpha)} \bigg(M_{\rm{HSE}} - \frac{P_{\rm{th}}r^2}{(1-\alpha)\rho G} \frac{{\rm{d}} \alpha}{{\rm{d}} r}\bigg), 
\label{eq: tot_mass_all}
\end{equation}
%---------------------
where $P_{\rm{th}}$ is the thermal pressure. Generally, $\alpha$ is expected to have some radial dependence. However, since we focus on a relatively small region (about 205--255 arcmin) in the outskirts, we hence assume that $\alpha$ is constant in this region. The gas mass fraction can then take the form:
%-------------------
\begin{equation}
f_{\rm{gas}}=\frac{M_{\rm{gas}}}{M_{\rm{tot}}}=f_{\rm{gas,HSE}}(1-\alpha),
\label{eq: fgas_all}
\end{equation}
%----------------------
suggesting that the gas mass fraction can be reduced by a factor of $1-\alpha$ in the presence of non-thermal pressure. To match our clumping-corrected gas mass fraction in the radial range of about 205--255 arcmin with the cosmic value, we therefore require $\alpha=0.21$, implying that 21 per cent of the total pressure is in the form of non-thermal pressure, agreeing with the predictions of numerical simulations \citep[e.g.][]{nelson2014weighing}.

\subsection{Correcting the gas entropy}
The gas entropy is of particular interest since it provides an insight into the thermal history of the ICM. Early X-ray observations \citep[see][and references therein]{Walker2019} found that the entropy measurements in the 0.6--1.0$r_{200}$ radial range are below the expected baseline entropy predicted by non-radiative simulations \citep{Voit2005}, indicating that non-gravitational process in the outskirts must be acting to reduce the measured entropy level. As indicated by numerical simulations \citep[e.g.][]{Nagai2011}, one possible factor for the relatively shallow entropy profile observed in the cluster outskirts is gas clumping. Clumps of dense gas would act to bias the observed surface brightness, causing the density to be overestimated, and hence the entropy to be underestimated.  

\begin{figure}
	\includegraphics[width=\columnwidth]{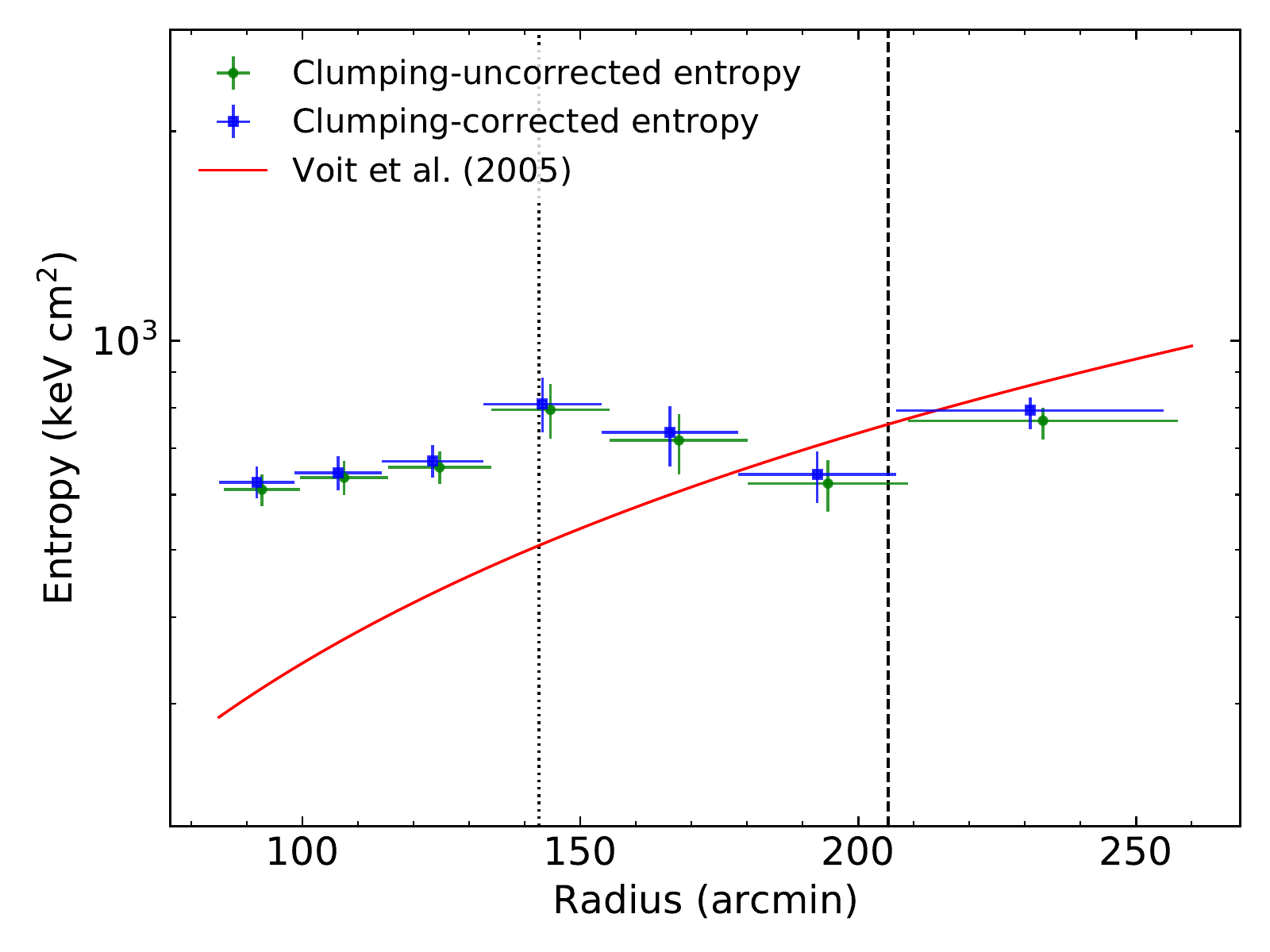}
	\caption{Measurements of the gas entropy along the the northern strip of the Virgo cluster with and without the clumping correction. The red solid line marks the power-law entropy predicted by non-radiative simulations \citep{Voit2005}. Beyond 170 arcmin, the clumping-corrected entropy measurements agree better with the baseline entropy profile predicted by non-radiative simulations.}
	\label{fig: entropy_N}
\end{figure}

In Fig. \ref{fig: entropy_N}, we show the measurements of the gas entropy along the the northern strip of the Virgo cluster with and without the clumping correction. In the same figure, we also show the power-law entropy profile predicted by non-radiative simulations for purely gravitational hierarchical structure formation. Beyond 170 arcmin, our clumping-corrected entropy measurements agree better with the baseline entropy profile predicted by non-radiative simulations. However, they are still lying below the baseline level, suggesting that clumping may not be the only factor that contributes to the entropy deficit. Below this radius, the clumping-corrected entropy measurements are in excess of the baseline level predicted by \citet{Voit2005}, as typically found for low-mass systems \citep{Pratt2010}, suggesting that the central AGN may be able to increase the gas entropy in the ICM above the baseline level out to around 170 arcmin.

%----------------------
%----------------------

\section{Discussion}
\subsection{Comparison with other measurements}
\label{sec: comparison_coma}
\begin{figure*}
	\includegraphics[width=0.9\textwidth]{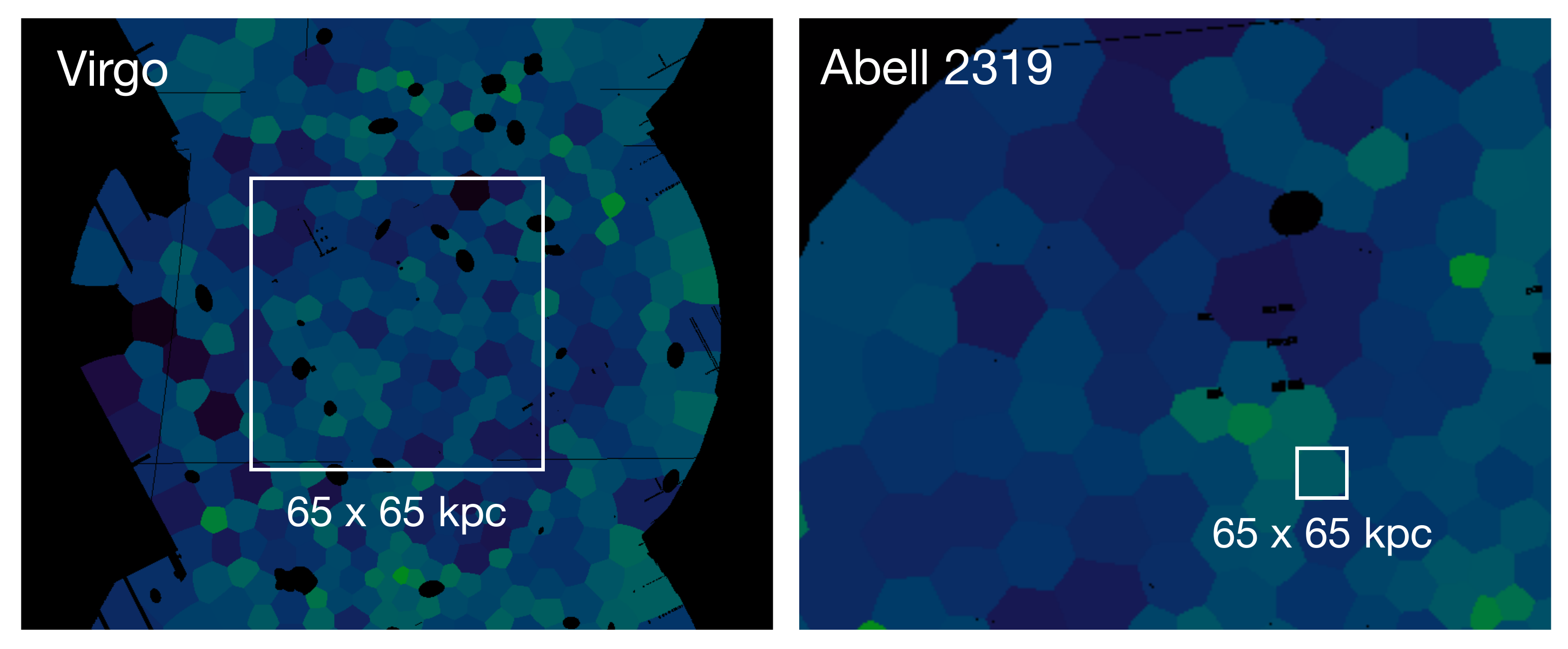}
	\caption{Comparing the spatial scale of the Voronoi tessellations of a 20 ks Virgo cluster observation around $r_{200}$ with the same depth observation of Abell 2319. Each tessellated region contains at least 20 counts. The white box shows the same spatial scale of $65 \times 65$ kpc for both clusters. Due to Virgo’s close proximity, the effects of gas clumping can be explored on very fine scales, about two orders of magnitude finer than can be achieved in Abell 2319. }
	\label{fig: A2319_Virgo}
\end{figure*}

In the current work, we explored the clumping level in the Virgo cluster on very fine scales, about two orders of magnitude finer than that achieved by exploring intermediate redshift clusters. In Fig. \ref{fig: A2319_Virgo}, we compare the spatial scale of the Voronoi tessellations of a 20 ks Virgo cluster observation around $r_{200}$ with the same depth observation of Abell 2319, a hot and massive cluster at $z = 0.0557$, where each tessellated region contains at least 20 counts. The white box shows the same spatial scale of $65 \times 65$ kpc for both clusters. The figure clearly shows that the level of gas clumping in the ICM of the Virgo cluster is explored on finer scales compared to Abell 2319.

%----------------------
%----------------------

\begin{figure}
	\includegraphics[width=\columnwidth]{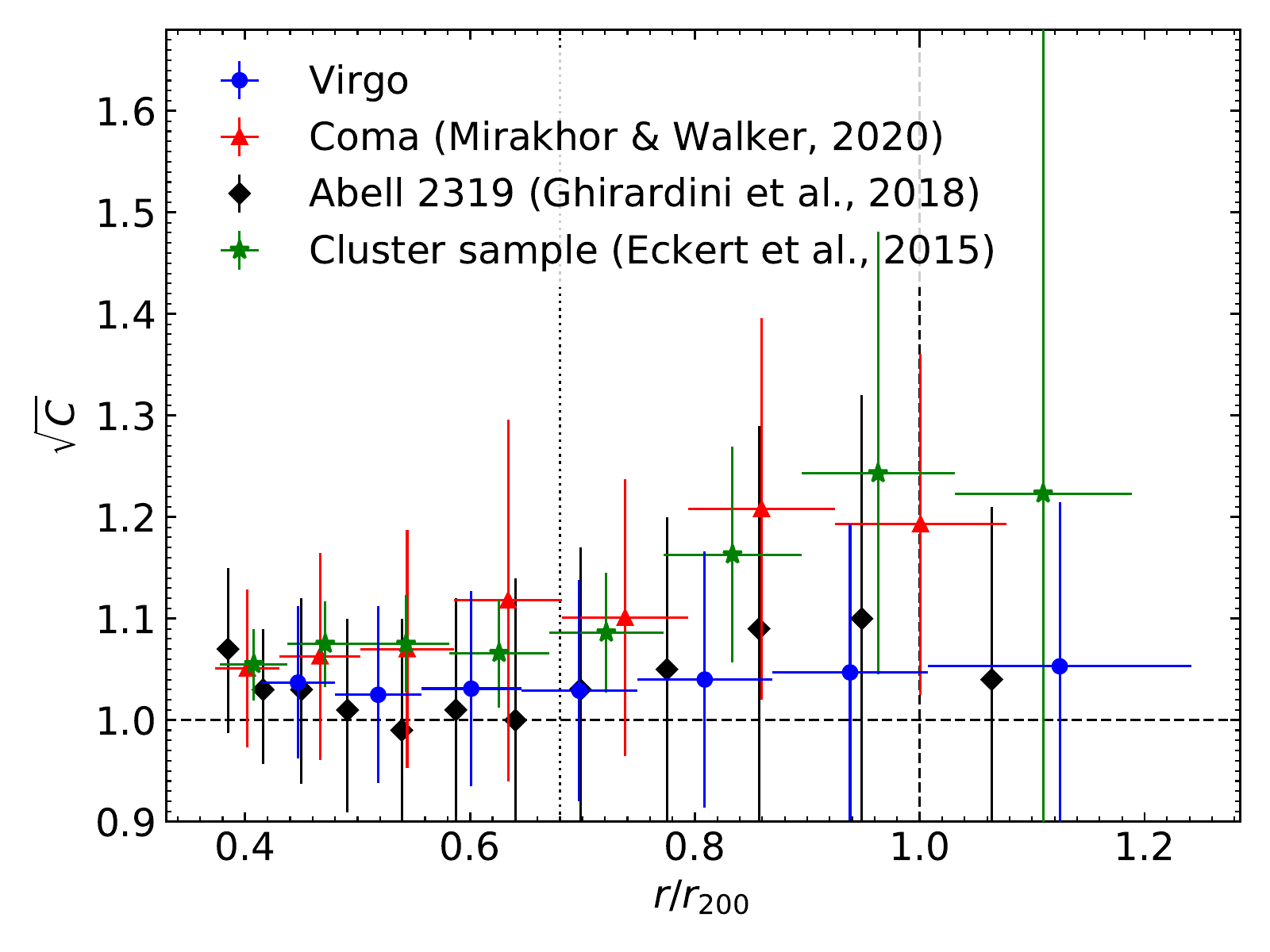}
	\caption{Comparison between the measurements of the clumping factor along the northern strip of Virgo (blue circle points) and the clumping factors recovered from previous measurements. The green star points mark the clumping factor estimated for a sample of 31 galaxy clusters in the redshift range 0.04--0.2 observed with \textit{ROSAT} \citep{eckert2015gas}. The red triangle and black diamond points represent the clumping factor estimated for Coma \citep{mirakhor2020complete} and Abell 2319 \citep{ghirardini2018xmm}, respectively. The vertical dotted and dashed lines represent the $r_{500}$ and $r_{200}$ radii, respectively. }
	\label{fig: comparison_observtion}
\end{figure}

As shown in Section \ref{sec: results}, the recovered clumping factors along the northern strip and other directions in the outskirts of Virgo are very low ($\sqrt{C} < 1.1$) out to the outskirts. These results indicate that the level of gas clumping is very mild in the ICM of the Virgo cluster, although the statistical uncertainty on the measurements is relatively large to reach a definite conclusion. These measurements are statistically in agreement with the clumping factor recovered for a sample of 31 galaxy clusters in the redshift range 0.04--0.2 imaged by \textit{ROSAT} \citep{eckert2015gas}, despite their measurements show a larger trend of increasing bias with radius (see Fig. \ref{fig: comparison_observtion}).

%----------------------
%----------------------

%----------------------
%\begin{figure*}
	%\includegraphics[width=0.9\textwidth]{histogram_Coma_A2319_Virgo_plot_v2.pdf}
	%\caption{Comparing the distribution of the X-ray surface brightness in the radial range 0.7--1.0$r_{200}$ of the massive Coma (left-hand panel) and Abell 2319 (right-hand panel) clusters with the surface brightness distribution along the northern strip of the Virgo cluster. The clumping factor recovered in this radial range of Coma is relatively higher than that recovered for the northern strip of Virgo, and this is clearly indicated by more asymmetric distribution of the surface brightness of Coma than Virgo.   }
	%\label{fig: histogram_Coma_Virgo}
%\end{figure*}
%----------------------

We also compared our clumping factor measurements along the northern strip with Abell 2319. \citet{ghirardini2018xmm} estimated the level of gas clumping present in the ICM of Abell 2319 by computing the ratio of the deprojected X-ray surface brightness profiles obtained from the mean and median of the azimuthal distribution, as is done in the current work. The recovered clumping factors in the ICM of this hot cluster are low too ($\sqrt C < 1.1$), and show a mild dependence on the radius, as shown in Fig. \ref{fig: comparison_observtion}. Their results are consistent with the recovered clumping factors for the northern strip of the Virgo cluster.

More recently, \citet{mirakhor2020complete} have presented a detailed study on the nearby Coma cluster using \textit{XMM--Newton} and \textit{Planck} data, and found that the clumping factor measurements increase slightly with radius to reach around 1.2 at $r_{200}$, as shown in Fig. \ref{fig: comparison_observtion}. The estimated clumping factor in the Virgo cluster might indicate that gas clumping is less significant in less massive systems ($kT \sim 4$ keV). This is consistent with some previous studies of lower-mass systems, such as the merging cluster Abell 1750 \citep{bulbul2016probing} and the fossil cluster RX J1159+5531 \citep{su2015entire}, but not all \citep[e.g. the galaxy group NGC 2563,][]{morandi2017gas}.

\subsection{Comparison with numerical simulations}
\label{sec: comparison_other}
To test the level of gas accretion with that predicted in the $\Lambda$ cold dark matter paradigm, we compared the clumping factor measurements along the northern strip of the Virgo cluster with the clumping factor profiles recovered from several sets of hydrodynamical cluster simulations. \citet{Nagai2011} analysed a sample of 16 simulated groups and galaxy clusters using hydrodynamical simulations, and found that the median clumping factor increases slightly within $r_{200}$ for the runs that include the cooling and star formation effects, which is statistically in agreement with our measurements. In non-radiative runs, however, they found that the median clumping factor is around 1.3 at $r_{200}$, which is significantly larger than that obtained in this work. Similar results are found when we compared our measurements with the predictions of other sets of numerical simulations \citep[e.g.][]{vazza2013properties,roncarelli2013large}. \citet{eckert2015gas} also found that non-radiative simulations systematically overestimate the level of gas clumping in the outskirts of galaxy clusters, while the runs that include radiative cooling and star formation provide a better match to the observations.

\begin{figure}
	\includegraphics[width=\columnwidth]{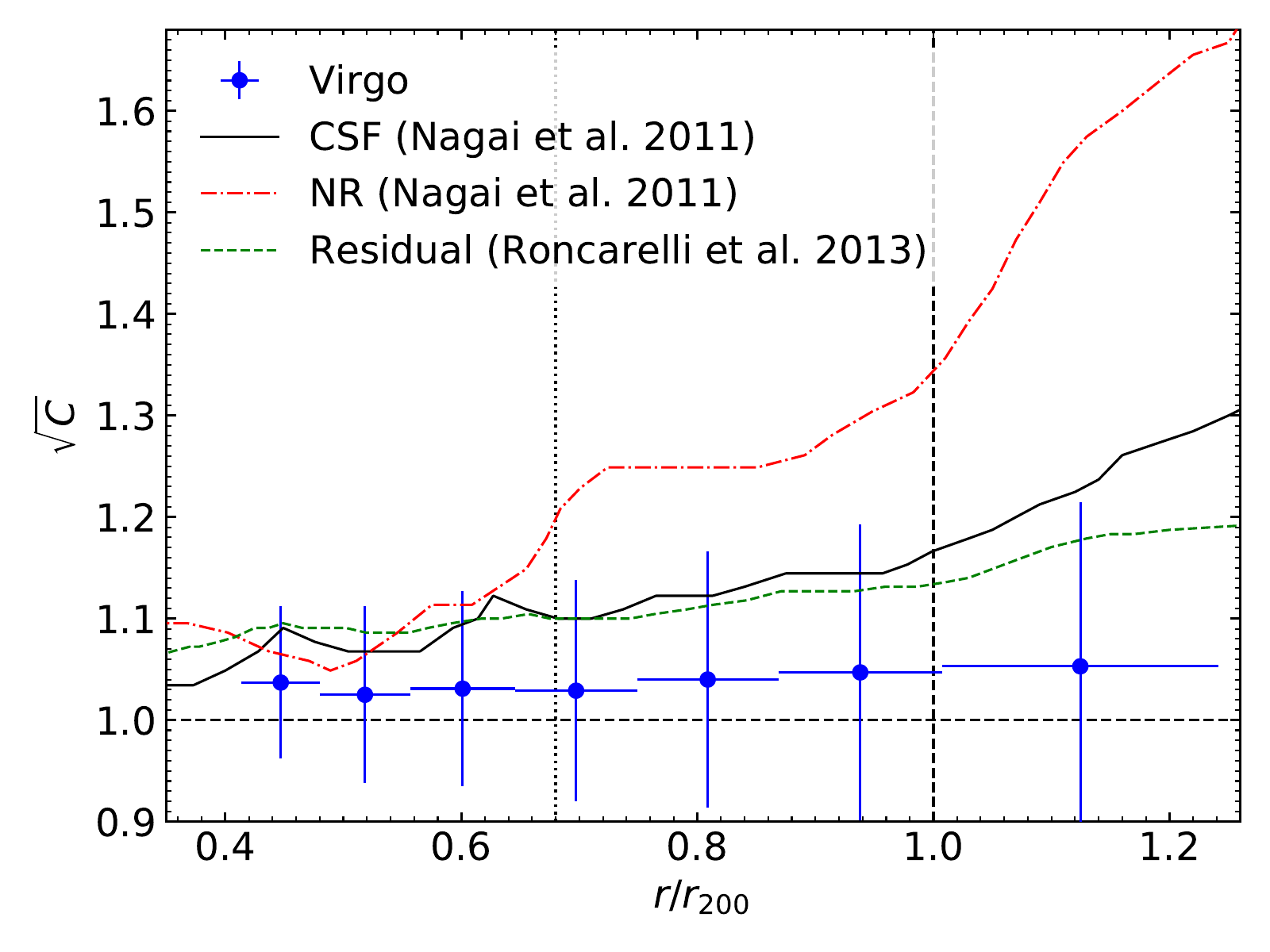}
	\caption{Comparison between the measurements of the clumping factor along the northern strip of Virgo (blue circle points) and the clumping factor profiles predicted in hydrodynamical cluster simulations \citep{Nagai2011}. The black solid and red dash-dotted lines represent the median radial profiles of the clumping factor for a sample of 16 simulated clusters in the gas cooling and star formation (CSF) and non-radiative (NR) runs, respectively. The green dashed line represents the "residual clumping" which is obtained by masking small dense clumps in the NR runs \citep{roncarelli2013large}. The vertical dotted and dashed lines represent the $r_{500}$ and $r_{200}$ radii, respectively.   }
	\label{fig: comparison_simulation}
\end{figure}

In Fig. \ref{fig: comparison_simulation}, we show a comparison between our clumping factor measurements and the clumping factor profiles predicted in non-radiative simulations and the simulations that include cooling and star formation \citep{Nagai2011}. In the same figure, we also compare the clumping factor measurements with the "residual clumping" \citep{roncarelli2013large}, which quantifies the inhomogeneity of the ICM induced by the large-scale accretion pattern without considering small dense clumps. As we can see in this figure, the hydrodynamical cluster simulations that include the physics of galaxy formation, such as radiative cooling, star formation, and supernova feedback, provide a good match to our measurements out to the outskirts. In contrast, the clumping factor profile predicted by the simulations that include only gravitational and hydrodynamical effects is considerably larger than our measurements, which might indicate that the level of gas clumping is overestimated in non-radiative simulations.

Another possible explanation for such difference between non-radiative simulations and our measurements is that our point-source detection procedure may have been able to detect most of the point sources and the prominent substructures in the field of Virgo. This would lower the level of gas clumping and explain the good match with the residual clumping profile, which is obtained by removing small dense clumps in the non-radiative simulations (the green dashed line in Fig. \ref{fig: comparison_simulation}). 

In addition to the effects of galaxy formation on gas clumping, the comparison between observed and simulated data also depends on the choice of sample. \citet{Nagai2011}, for instance, examined the dependence of gas clumping on the cluster mass, and found that lower-mass clusters have a smaller clumping factor. They argued that this is because these systems, compared with massive clusters, have a larger fraction of lower-temperature gas that is not detectable in X-rays. Also, it is found that perturbed systems are, on average, characterised by a larger value of the gas clumping factor at all radii \citep[e.g.][]{vazza2013properties}. This enhancement is mainly due to the massive accretion of gas along filaments, which is more active in perturbed systems than in relaxed systems. It is worth noting that the simulated sample used in this work have a mix of disturbed and relaxed clusters with a wide range of masses, and we therefore comparing the average clumping factor of the simulated sample with our clumping factor measurements. %Thus, any differences between our measurements and numerical simulations is unlikely due to a sample choice. 

%----------------------
%----------------------

\section{Conclusions}
In this work, we have investigated in detail gas clumping down to scales of $5 \times 5$ kpc using \textit{XMM--Newton} observations of the Virgo cluster outskirts, providing the first high-resolution exploration of gas clumping. We have also investigated gas clumping around the virial radius of Virgo using multiple observations in the region to derive the azimuthal variation of the clumping factor, and to test whether it is greatest along the north-south filament direction. Our results can be summarized as follows:

\begin{enumerate}
\item By masking point sources and extended substructures in the Virgo field, and tessellating the resulting images into Voronoi tessellations, we have found that the clumping factor profile along the northern strip of the Virgo cluster recovered by dividing the mean of the deprojected X-ray surface brightness profile by the median of the same profile in a given annulus is very mild ($\sqrt{C}$ < 1.1), and shows a slight trend of increasing with increasing radius. This indicates that the bias introduced by unresolved gas clumping is very low in the northern strip of Virgo, although the statistical uncertainty on the measurements is relatively large to reach a definite conclusion.

\item We have found that the unresolved clumping factor measurements on small scales are highly uniform in all directions, and are not enhanced along the north-south direction. However, this finding is based on investigating of a small fraction of the Virgo outskirts and, therefore, we cannot rule out the possibility of an asymmetric distribution of the clumping factor if the entire outskirts is considered. Alternatively, it is still possible that gas clumping is enhanced along the north-south direction, but our point-source detection procedure may have been able to detect most of the point sources and the prominent substructures in the field down to scales of $5 \times 5$ kpc. This would explain the moderate level of unresolved gas clumping, and the good agreement among the measurements of the clumping factors seen along different directions in the Virgo outskirts.

\item When accounting for the bias introduced by gas clumping, the gas mass fraction along the northern strip drops by about 5 per cent, agreeing with the universal hot gas mass fraction at $r_{200}$. In the radial range of about 205--255 arcmin (corresponding roughly to [1.0--1.2]$r_{200}$), however, our clumping-corrected gas mass fraction is still higher than the universal value, which may indicate a non-negligible contribution from non-thermal pressure in the region. In this radial range, we have found that a support of about 20 per cent from the non-thermal to total pressure is required to reproduce the expected value of the universal gas mass fraction in the outskirts of the northern strip of the Virgo cluster, agreeing with the predictions of numerical simulations.

\end{enumerate}

\section*{Acknowledgements}
We thank the referee for their helpful report. MSM and SAW acknowledge support from the NASA \textit{XMM--Newton} grant 19-XMMNC18-0030. Based on observations obtained with \textit{XMM--Newton}, an ESA science mission with instruments and contributions directly funded by ESA Member States and NASA.

%%%%%%%%%%%%%%%%%%%%%%%%%%%%%%%%%%%%%%%%%%%%%%%%%%
\section*{Data Availability}
The \textit{XMM--Newton} Science Archive (XSA) stores the archival data used in this paper, from which the data are publicly available for download. The \textit{XMM} data were processed using the \textit{XMM--Newton} Science Analysis System (SAS). The software packages \textsc{heasoft} and \textsc{xspec} were used, and these can be downloaded from the High Energy Astrophysics Science Archive Research Centre (HEASARC) software web-page. Analysis and figures were produced using \textsc{python} version 3.7.



\bibliographystyle{mnras}
\bibliography{Virgo} % if your bibtex file is called example.bib

%\begin{thebibliography}{99}
%\end{thebibliography}

%%%%%%%%%%%%%%%%%%%%%%%%%%%%%%%%%%%%%%%%%%%%%%%%%%

%%%%%%%%%%%%%%%%% APPENDICES %%%%%%%%%%%%%%%%%%%%%

\appendix

%%%%%%%%%%%%%%%%%%%%%%%%%%%%%%%%%%%%%%%%%%%%%%%%%%

% Don't change these lines
\bsp	% typesetting comment
\label{lastpage}
\end{document}